\documentclass[acmsmall,screen]{acmart}
\AtBeginDocument{%
  }

\setcopyright{acmlicensed}
\copyrightyear{2018}
\acmYear{2018}
\acmDOI{XXXXXXX.XXXXXXX}
\acmConference[Conference acronym 'XX]{Make sure to enter the correct
  conference title from your rights confirmation email}{June 03--05,
  2018}{Woodstock, NY}
\acmISBN{978-1-4503-XXXX-X/2018/06}

\usepackage{pifont}
\usepackage{enumitem}
\usepackage{multirow}
\usepackage{colortbl}
\usepackage{wrapfig}
\usepackage{framed}

\theoremstyle{definition}
\newtheorem{definition}{Definition}[section]

\definecolor{shadecolor}{gray}{0.93}
\newenvironment{custommdframed}
{%
    \vspace{-0.2cm}%
    \MakeFramed{\advance\hsize-6pt\FrameRestore}%
    \noindent%
}
{\endMakeFramed}

\definecolor{DeepRed}{RGB}{178,34,34}
\definecolor{LightGray}{gray}{0.92}
\definecolor{DeepGray}{RGB}{64,64,64}

\newcommand{\ourbench}{\textsc{LoCoEval}}

\begin{document}

\title{A Scalable Benchmark for Repository-Oriented Long-Horizon Conversational Context Management}

\author{Yang Liu}
\email{liuyang26@buaa.edu.cn}

\author{Li Zhang}
\email{lily@buaa.edu.cn}

\author{Fang Liu}
\authornote{Corresponding author}
\email{fangliu@buaa.edu.cn}

\author{Ping Lin}
\email{linping@buaa.edu.cn}

\author{Xinyi Li}
\email{lixinyi2023@buaa.edu.cn}

\affiliation{%
  \institution{State Key Laboratory of Complex \& Critical Software Environment, School of Computer Science and Engineering, Beihang University}
  \city{Beijing}
  \country{China}
}

\renewcommand{\shortauthors}{Yang Liu et al.}

\begin{abstract}
In recent years, large language models (LLMs) have advanced rapidly, substantially enhancing their code understanding and generation capabilities and giving rise to powerful code assistants.
However, in practical repository development, excessively long-horizon conversational context may overwhelm models, causing the loss of critical information and degraded performance, thereby limiting the utility of code assistants. Existing context management methods proposed to mitigate this context dilemma primarily target general-purpose conversations, while repository-oriented solutions remain largely unexplored, which is largely due to the lack of reliable evaluation benchmarks. To bridge this gap, we present \ourbench{}, the first long-horizon conversational context management benchmark tailored to repository-oriented development scenarios. Adhering to three key principles, \ourbench{} is constructed via an LLM-driven pipeline that generates realistic and diverse repository-oriented conversations, capturing key interaction patterns such as iterative requirements, noisy input, and retrospective questions. It features 128 samples divided into two subsets (\textit{i.e.}, single-hop and multi-hop) and supports three types of evaluation tasks.
Each sample contains an average of 2.5 requirements and 50 conversational turns, resulting in total context lengths reaching up to 64K\textasciitilde256K tokens.
We evaluate 7 baselines, including 4 representative context management methods, using 3 advanced backbone LLMs on \ourbench{}. The results reveal substantial challenges faced by standalone LLMs and existing approaches, especially memory systems, in repository-oriented conversational scenarios.
To address these limitations, we further propose an improved method integrating conversational and repository information into a unified memory, which outperforms all baselines (\textit{Oracle} excluded) and demonstrates robustness. 
Additionally, we investigated the impact of various factors on method performance, providing actionable insights for future research. Our benchmark and improved method are available at \url{https://anonymous.4open.science/r/LoCoEval}.
\end{abstract}



\maketitle

\section{Introduction}

In recent years, the rapid development of large language models (LLMs) and the significant improvement of their coding capabilities have given rise to a variety of advanced LLM-based code assistants, such as 
Cursor \cite{Cursor}, Claude Code \cite{ClaudeCode}, and Trae \cite{Trae}, which demonstrate strong performance across a wide range of programming tasks. In their most widely used repository-oriented development scenarios, these assistants can interact with developers through natural-language conversations, while accessing files within the repository and invoking various tools to automatically understand the codebase, implement requirements, or fix bugs, thereby substantially improving development efficiency and productivity. 
However, due to the inherent complexity of real-world repository development, a single requirement often undergoes multiple iterations. Moreover, multiple requirements may progress in parallel, with additional clarification questions interleaved throughout the process. As a result, a single conversation can easily span dozens or even hundreds of turns. As the number of turns increases, the amount of context that the LLM needs to process grows accordingly. Excessively long contexts may lead to critical information being truncated once the context window is exceeded and can also disrupt model reasoning \cite{yang2025llm, wang2025breaking}, reducing response accuracy and consistency while incurring unnecessary computational overhead. Together, these issues limit the applicability of code assistants in broader and more complex development scenarios.

\begin{wrapfigure}{r}{0.4\textwidth}
    \centering
    \setlength{\abovecaptionskip}{0.1cm}
    \includegraphics[width=\linewidth]{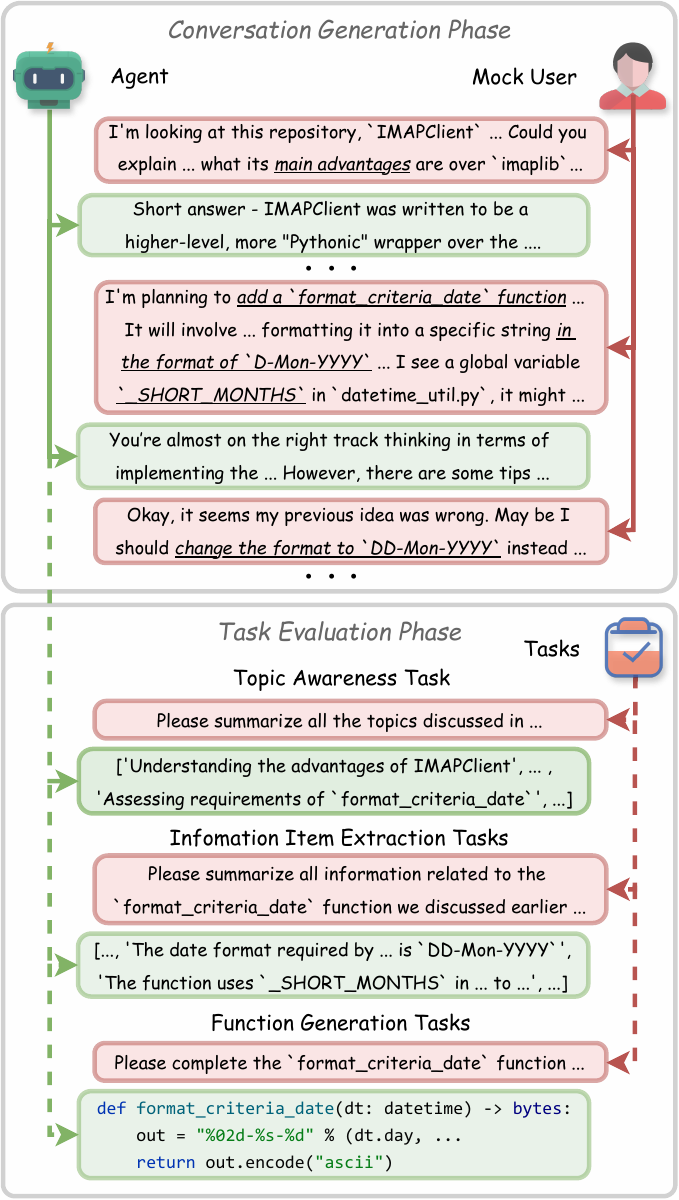}
    \caption{An example of \ourbench{}. All mock user queries and agent responses are dynamically generated during evaluation.}
    \vspace{-0.2cm}
    \label{fig:example}
\end{wrapfigure}

To address the challenges posed by excessive context in long-horizon conversations, researchers have conducted extensive studies on \textit{context management}, aiming to streamline LLM context through efficient organization and retrieval of contextual information. Early work primarily focused on static \textit{context compression}. More recently, with the rapid advancement of LLM's agentic capabilities \cite{comanici2025gemini, team2025kimi, acharya2025agentic}, increasing attention has been paid to dynamic \textit{memory system}, which enables more accurate handling of fine-grained or complexly interrelated information. 
Context management agents built on these methods can interact with users in the same manner as a standalone LLM, while using less context and improving consistency. 
However, existing methods are largely designed for general-purpose conversations and have not been specifically optimized for repository-oriented scenarios. The key particularity of such scenarios is that the contextual information originates not only from the conversation but also from textual and code artifacts within the repository, with the two sources often being tightly intertwined. Consequently, effective context management in this scenario may require mechanisms that can efficiently integrate both sources of context, which current methods lack. At present, the primary bottleneck hindering progress in this direction is the absence of a reliable benchmark for long-horizon, repository-oriented conversational context management, which in turn prevents objective and systematic performance evaluation.

To this end, we propose \ourbench{}, the first long-horizon conversational context management benchmark for repository-oriented scenarios, aiming to evaluate to what extent context management methods can accurately retain and recall critical contextual information while compressing the context in repository-oriented long-horizon conversations, thereby improving answer correctness. To construct \ourbench{}, we design a novel LLM-driven automated pipeline that generates query outlines from samples in existing repository-level code generation datasets. These outlines are used to generate mock\footnote{The term ``mock'' is used here because queries are generated by an LLM acting as a user, rather than by real human users.} user queries that interact with context management agents, and thus define the overall trajectory of the conversation. At the end of the conversation, a set of predefined task queries is issued, requiring the agent to accurately recall and leverage information from earlier queries while effectively combining it with information from the repository to produce correct answers. The accuracy of the agent's responses thus reflects its context management capability. Figure \ref{fig:example} illustrates an example of such a conversation. 
Given that incorrect queries may propagate biased information and consequently lead to erroneous evaluations, while a lack of realism and diversity in conversations may cause evaluations not broadly representative or deviate from real-world scenarios, we argue that \ourbench{} needs to satisfy three key principles of \ding{182} correctness, \ding{183} realism, and \ding{184} diversity, 
while aligning with the target scenario.
Adhering to these principles, the generated conversations exhibit the following characteristics: \ding{172} queries reflect key real-world interaction patterns, including iterative requirements, noisy content, and retrospective questions; \ding{173} both user queries and agent responses are generated dynamically during evaluation; and \ding{174} each conversation contains 1\textasciitilde4 preset requirements
and reaches a total context length 
of approximately 64K\textasciitilde256K tokens.
The resulting \ourbench{} is divided into two subsets according to concentrated or dispersed information distribution patterns, each containing 64 benchmark samples. Each sample consists of one query outline and three types of evaluation tasks, \textit{i.e.}, topic awareness, information item extraction, and function generation. 
More statistics are presented in Table \ref{tab:statistics}.

We conduct a comprehensive evaluation of 7 representative baselines on \ourbench{} using 3 advanced backbone LLMs. The results show that, despite with preliminary adaptation via an RAG strategy, both standalone LLMs and existing general-purpose context management methods still face substantial challenges in handling repository-oriented long-horizon conversations, often failing to effectively leverage repository information. This limitation is particularly evident for memory systems, the current mainstream paradigm. To address this, we develop an improved variant of Mem0 \cite{chhikara2025mem0}, a representative general-purpose memory system, termed Mem0$^\mathcal{R}$. By enabling preliminary integration of conversation history and code repository, Mem0$^\mathcal{R}$ outperforms the strongest baseline agent. Finally, we analyze the impact of various external factors on context management performance, with the aim of providing useful insights for future research. Overall, the contributions of this work are summarized as follows:

\begin{itemize}[leftmargin=*]
    \item We propose \ourbench{}, the first benchmark specifically focusing on evaluating long-horizon conversational context management in repository-oriented scenarios. We design a novel automated pipeline that constructs \ourbench{} based on existing datasets. Through multiple strategies, \ourbench{} achieves substantial correctness, realism, and diversity, closely mirroring real-world workflows in which developers interact with code assistants during repository development.
    \item We conduct a comprehensive evaluation of representative context management methods and backbone LLMs on \ourbench{}, accompanied by in-depth analysis. The results reveal substantial challenges that existing methods face when handling repository-oriented long-horizon conversations, providing valuable insights for future research.
    \item Based on detailed analysis, we develop a repository-specific context management method, Mem0$^{\mathcal{R}}$, through a simple yet effective extension of Mem0. By integrating conversational and repository information into memory and enabling context-aware repository retrieval, Mem0$^{\mathcal{R}}$ achieves the best overall performance.
\end{itemize}

\section{Related Work}

\subsection{Conversational Context Management for LLMs}
While LLMs excel at multi-turn interactions and complex tasks, long-horizon conversations lead to growing context that exceeds the model's finite context window, causing information loss and degraded reasoning \cite{yang2025llm, wang2025breaking}.
To address this challenge, existing work has explored two main directions: internal model architecture improvements and external context management. For the former, mainstream LLMs \cite{liu2024deepseek, comanici2025gemini, anthropic2024claude, yang2025qwen2, openai2025gpt5mini} have continuously expanded their context windows and optimized long-context capabilities through techniques such as sparse attention \cite{yuan2025native, jiang2024minference}, relative position encoding \cite{foumani2024improving}, and RoPE \cite{liu2023scaling}. Nevertheless, these approaches do not fundamentally resolve the contradiction between a finite context window and unbounded context length. Consequently, researchers have increasingly focused on context management solutions.  

Context management aims to effectively organize and retrieve contextual information, providing LLMs with more concise background knowledge under a bounded length. Early studies primarily adopted context compression strategies, selecting and retaining useful parts of the context by modifying the model \cite{ge2023context, zhang2024long, jiang2024longllmlingua} or applying RAG \cite{xu2024recomp, rau2024context, csakar2025maximizing}. However, such one-shot processing lacks scalability and struggle to handle fine-grained or complexly interrelated contextual information. This has driven researchers to turn toward memory systems \cite{packer2023memgpt, zhong2024memorybank, li2025memos, tan2025prospect, chhikara2025mem0, xu2025mem} as a more advanced approach to context management. Inspired by human memory mechanisms, these methods leverage structured memory databases to extract, integrate, and retrieve important information from conversational context, enabling more effective context management. 
As a pioneering work, MemGPT \cite{packer2023memgpt} implemented a memory system referred to as virtual context management.
To further improve memory quality and retrieval adaptability, subsequent studies have explored various directions.  
For example, Mem0 \cite{chhikara2025mem0} maintains memory through a database-like manner, whereas Agentic Memory \cite{xu2025mem} organizes memories in a graph structure to establish semantic links among them.
Notably, existing context management methods are primarily designed for general LLM conversations, and there are currently no approaches specifically optimized for repository-oriented scenarios.

As the output of these context management approaches, the retrieved contextual information is integrated into the user query and passed to an LLM to generate a response. Together, the context management method and the response-generating LLM form an independent interactive system, which we refer to as a \textbf{context management agent}. From the user's perspective, its interaction pattern is identical to a standalone LLM's question-answer loop, while reducing context overhead and improving response accuracy and consistency. Such agents constitute the evaluation targets of context management benchmarks, including our \ourbench{}.

\begin{table}[t]
\centering
\setlength{\abovecaptionskip}{0.1cm}
\small
\caption{Statistics of \ourbench{}.}
\label{tab:statistics}
\begin{tabular}{ccccccc}
\toprule
\textbf{Subsets} & \textbf{Samples} & \textbf{Tasks} & \textbf{Repos} & \textbf{Requirements/sample} & \textbf{Turns/sample} & \textbf{Tokens/sample} \\
\midrule
2 & 128 & 768 & 37 & 1\textasciitilde4 & 30\textasciitilde70 & 64K\textasciitilde256K \\
\bottomrule
\end{tabular}
\vspace{-0.3cm}
\end{table}

\subsection{Conversational Context Management Benchmark}

To evaluate context management of various methods, a number of benchmarks have been proposed \cite{maharana2024evaluating, wu2024longmemeval, chen2025halumem, jiang2025personamem, hu2025evaluating, deshpande2025memtrack}. For instance, LoCoMo \cite{maharana2024evaluating} 
evaluates long-range context management through tasks such as question answering and event summarization, 
revealing that current LLMs still struggle with long-range temporal and causal understanding. LongMemEval \cite{wu2024longmemeval} is designed to assess long-horizon memory in chat assistants across five abilities, including information extraction, temporal reasoning, and knowledge updating. 
PersonaMem-v2 \cite{jiang2025personamem} evaluates long-context personalization by simulating realistic user-assistant conversations,
highlighting that reasoning
is the main bottleneck.
However, 
there are currently no benchmarks that focus on evaluating context management in repository-oriented scenarios. This significantly hinders the development of related approaches and is precisely the problem we aim to address.

\subsection{Repository-Oriented Benchmark}

The coding capabilities of LLMs, particularly in repository-oriented development, have attracted substantial attention from both academia and industry.
In recent years, several benchmarks have been proposed to quantitatively evaluate repository-level development capabilities from different perspectives. Early benchmarks were limited to static settings, where the evaluation is performed in a single-shot manner and overlooks the examination of multi-turn interactions.
Representative tasks include single-turn repository-level code generation \cite{zhang2023repocoder, ding2023crosscodeeval, yu2024codereval, li2024deveval, li2024evocodebench, zhao2025towards}, code understanding \cite{li2025longcodeu, peng2025swe, qiu2025locobench, du2025dependeval}, code translation \cite{wang2024repotransbench, ou2024repository, zhang2025skeleton}, issue resolving \cite{jimenez2023swe, zan2025multi, zhang2025swe}, and vulnerability detection \& repair \cite{yildiz2025benchmarking, sunrepofixeval, liu2025securereviewer}. However, in real-world usage scenarios, interactions between users and LLMs-based systems are often multi-turn and dynamic, which further requires benchmarks to simulate such interaction patterns in order to evaluate the corresponding capabilities. Depending on the specific capabilities of interest, several multi-turn benchmarks are proposed, which can be roughly categorized into several groups: \ding{182} multi-turn instruction-following ability, such as CoCoPIF \cite{hancocopif} and CodeIF-Bench \cite{wang2025codeif}; \ding{183} multi-turn tool-use ability, such as LoCoBench-Agent \cite{qiu2025locobenchagent}; and \ding{184} iterative generation or question-answering ability, such as SR-Eval \cite{zhan2025sr}, CodeAssistBench \cite{kim2025codeassistbench}, and CodeFlowBench \cite{wang2025codeflowbench}. Nevertheless, the contexts provided by these benchmarks are generally short, and they have not yet focused on the long-horizon context management capability of LLM-based systems in repository-oriented multi-turn interactions. Such capability is crucial for maintaining accuracy and consistency over extended conversations, highlighting an important research gap.

\section{Approach}

\subsection{Automated Construction of \ourbench{}}
\label{sec:automated_construction_of_locoeval}

The essence of conversational context management lies in organizing, maintaining, and retrieving context information relevant to the target task (\textit{e.g.}, code generation) from the conversation
to enhance generation. Accordingly, to construct \ourbench{}, we adopt an inverse perspective.
Specifically, rather than starting from raw conversations,
we first identify the information that is useful for completing the current task, then embed it into the user queries 
to form a coherent conversation containing the necessary contextual cues. This synthesized conversation is subsequently used to evaluate context management capabilities of agents. In particular, considering the executability of programming languages, the primary requirements of repository-oriented scenarios, and the prevailing practices of existing repository-level benchmarks \cite{li2024deveval, yu2024codereval, jimenez2023swe, li2024evocodebench, yang2024execrepobench, wang2025coderag}, we choose the \textbf{function generation} evaluated by the Pass@k metric as the core task of \ourbench{}. 

\begin{figure}[t]
    \centering
    \setlength{\abovecaptionskip}{0.1cm}
    \includegraphics[width=\linewidth]{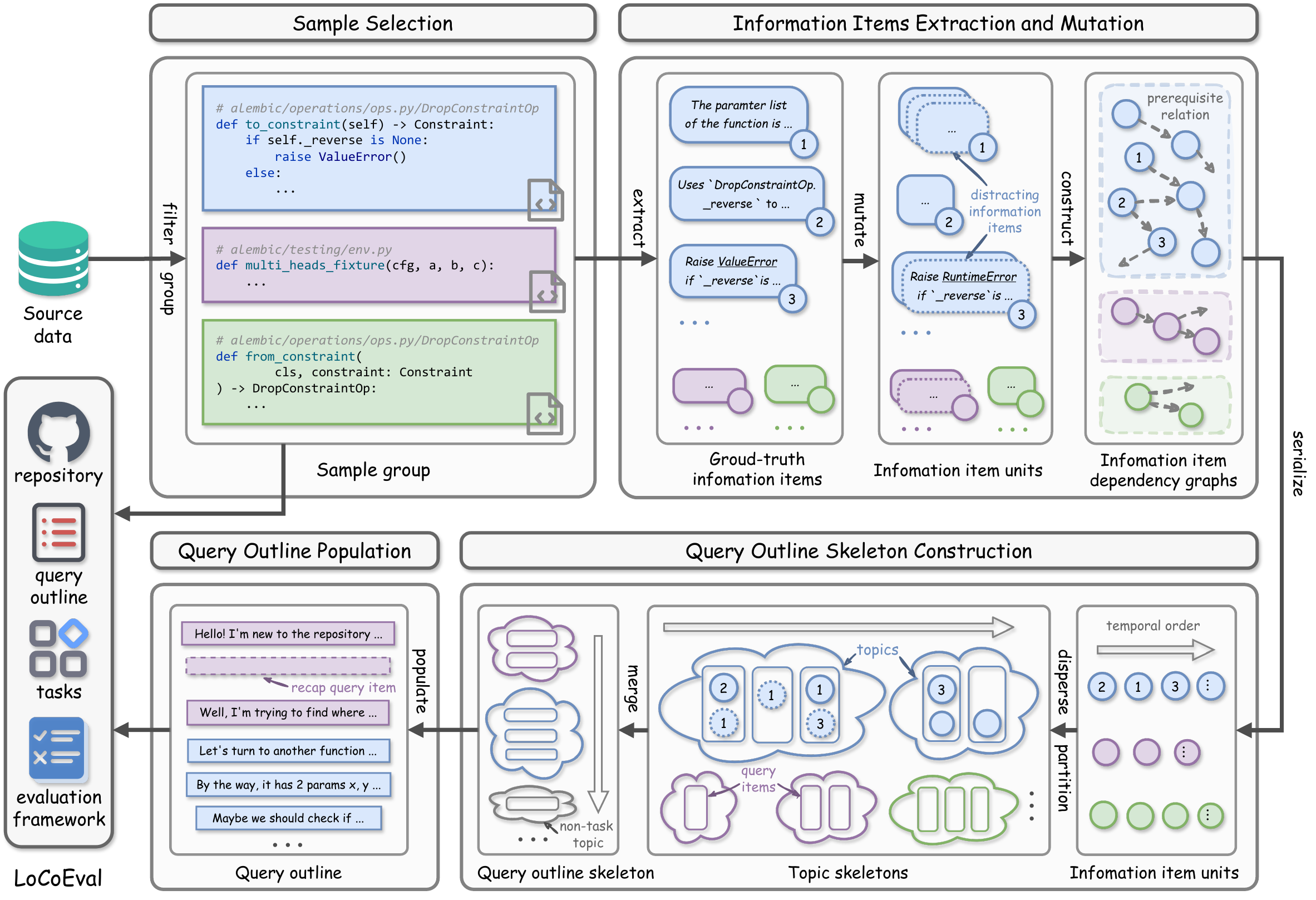}
    \caption{Overview of the construction pipeline of \ourbench{}.}
    \vspace{-0.4cm}
    \label{fig:construction_overview}
\end{figure}

\ourbench{} is built upon DevEval \cite{li2024deveval}, a repository-level function generation dataset containing 1,825 samples\footnote{From \S \ref{sec:automated_construction_of_locoeval} to \S \ref{sec:benchmark_components} (excluded), the term ``sample'' all refers to an instance from DevEval dataset.} collected from 117 repositories. An overview of the construction pipeline of \ourbench{} is shown in Figure \ref{fig:construction_overview}. First, suitable candidate samples are selected from DevEval (\S \ref{sec:sample_selection}). 
Next, information that is crucial for completing the target function is extracted from the target function's reference implementation of each sample, which is then mutated to obtain distracting information (\S \ref{sec:information_items_extraction_and_mutation}). Finally,
all information is dispersed into a unified query outline to construct its skeleton (\S \ref{sec:query_outline_skeleton_construction}), which is then populated using an LLM (\S \ref{sec:query_outline_population}). 
By applying a variety of strategies during construction, we ensure that the three key principles are amply adhered to. We will further systematically elaborate the adherence in \S \ref{sec:adherence_to_key_principles}.
All LLMs used here are \textit{Gemini 2.5 Pro} \cite{comanici2025gemini}.

The outline is the core component of \ourbench{}. It provides an appropriate structure and sufficient content to support the generation of mock user queries,
through which information about target functions is gradually conveyed to the agent, enabling the evaluation of conversational context management.
In addition to the outline, \ourbench{} includes several other components easy to obtain, which will be introduced in detail in \S \ref{sec:benchmark_components}.
Notably, both the mock user queries and agent responses are generated dynamically in the multi-turn interactions during evaluation. This design aligns with the way agents are used in the real world, thereby preserving realism and semantic coherence, and ensuring consistency in the source of agent responses within each conversation.

\subsubsection{Sample Selection}
\label{sec:sample_selection}

The core object of conversational context management is the conversation. In repository-oriented scenarios, however, relevant code in the repository can also independently provide valuable information for function generation. To ensure that information is primarily derived from the conversation rather than from repository alone, thereby enabling a more faithful evaluation of context management, we first perform a sample selection to filter out samples that can be solved by relying solely on repository code.
Specifically, we adopt a sliding-window RAG strategy. For each sample, we use the signature 
of the target function as the query to retrieve the top-10 most similar code snippets from the repository.
We then use one of the most powerful code LLMs, \textit{Claude Sonnet 4.5}
\cite{anthropic2024claude}, to generate a solution for each sample, and evaluate it using the corresponding test cases. Samples that pass all test cases are excluded, resulting in the removal of 788 samples in total. After this, the remaining samples generally cannot be solved correctly by relying solely on repository retrieval, making the conversation a critical source of information.

\subsubsection{Information Items Extraction and Mutation}
\label{sec:information_items_extraction_and_mutation}
Next, we use an LLM together with static analysis tools to extract information that is crucial for correctly completing the target function, and then perform controlled mutation to generate distracting information, which simulates the presence of noisy, incomplete, and potentially misleading cues commonly encountered during real-world repository development.
For clarity, in this subsection as well as in \S \ref{sec:query_outline_skeleton_construction} and \S \ref{sec:query_outline_population}, we first introduce several concept definitions, followed by a detailed description of our method.

\begin{definition}[ground-truth information item]
A ground-truth information item is an atomic piece of information extracted from the reference implementation of a target function that is crucial for its correct completion, such as a cross-file dependency.
Each ground-truth information item consists of a natural language \underline{\textit{description}} 
and, optionally, the path \underline{\textit{locations}} of the associated repository code elements.
\end{definition}

\begin{definition}[distracting information item]
A distracting information item is derived by mutating the corresponding ground-truth information item. It shares the same structure and the same or a similar semantic scope as the ground-truth one, 
while introducing semantic inconsistencies that conflict with the original information.
Ground-truth and distracting information items are collectively referred to as \textbf{information items}, which can be conveyed to the context management agent through conversational user queries. 
Table~\ref{tab:information_types} shows the types of information items.
\end{definition}

\begin{definition}[information item unit]
An information item unit consists of each ground-truth information item and its corresponding distracting information items. Formally, let ground-truth information item be denoted by $I$, and its associated distracting information items by $\tilde{I}_1, \ldots, \tilde{I}_M$. The resulting information item unit is defined as $U(I) = \{ I, \tilde{I}_1, \ldots, \tilde{I}_M \}$.
\end{definition}

\begin{definition}[prerequisite relation]
If an information item (unit) $B$ semantically depends on another information item (unit) $A$ as contextual support, then $A$ is considered a prerequisite information item (unit) for $B$, and such relation is defined as \textit{prerequisite relation}. For example, $A$ and $B$ could be ``The function parameter list is $(x, y)$'' and ``The function should first checks whether $y$ is zero'', respectively.
\end{definition}

\begin{definition}[information item dependency graph]
An information item dependency graph is a directed acyclic graph (DAG) formed by linking all information items (or item units) within a single sample via prerequisite relations, explicitly modeling the logical dependencies among them.
\end{definition}

In this step, we first extract all ground-truth information items 
of each sample. For the first four types (Type 1\textasciitilde4), we use static analysis tools for extraction, while the remaining five types (Type 5\textasciitilde9) are extracted using an LLM. 
For information items extracted by the LLM, we require the LLM to identify their prerequisite information items during extraction, 
and other prerequisite relations are determined based on predefined rules. As a result, all ground-truth information items are organized into an information item dependency graph.

\begin{table}[t]
\centering
\setlength{\abovecaptionskip}{0.1cm}
\caption{Types of Information Items.}
\label{tab:information_types}
\resizebox{\linewidth}{!}{
\begin{tabular}{lll}
\toprule
\textbf{ID} & \textbf{Name} & \textbf{Brief Description} \\
\midrule
1 & Function Location & Including the file path and optionally the containing class \\
2 & Internal Dependency & Function dependency on internal repository code elements \\
3 & External Dependency & Function dependency on external libraries, such as third-party libraries \\
4 & Parameter List & List of former parameters of the function \\
5 & Core Functionality & Core functionality or logic of the function \\
6 & Repo-Specific Knowledge & Knowledge specific to the repository and related to the function \\
7 & Logical Constraint & Specific handling logic of the function, such as edge case handling \\
8 & Coding Convention & Specific coding patterns that the function should follow \\
9 & Others & Other appropriate information \\
\bottomrule
\end{tabular}
}
\vspace{-0.2cm}
\end{table}

Then, for each sample, we randomly select half of the ground-truth information items for mutation. For the first three types (Type 1\textasciitilde3), which involve code locations, mutation is performed by replacing the original locations with similar ones identified elsewhere in the repository. For the remaining six types (Type 4\textasciitilde9), 
an LLM is used to modify the original descriptions. The resulting distracting information items are added to
the sample's information item dependency graph according to their counterparts, yielding a dependency graph composed of information item units. Each mutated unit contains 1\textasciitilde3 distracting information items, while the other units are left without any distracting items.

\subsubsection{Query Outline Skeleton Construction}
\label{sec:query_outline_skeleton_construction}

To enhance diversity and simulate realistic repository development scenarios, where developers often simultaneously handle multiple tasks within the same repository, we partition samples from the same repository into sample groups, each containing several samples. All information items within the same group are used to construct a single query outline, serving as the basis for dynamically generating mock user queries during evaluation.

\begin{definition}[query item]
A query item contains necessary information for generating a mock user query. Each query item corresponds to exactly one query and primarily includes:

\begin{itemize}[leftmargin=*]
  \item \textit{Query type}: the intent type of the query, aligned with the repository-level code question taxonomy proposed by~\citet{peng2025swe}.
  \item \textit{Contained information}: the information items
  that the query should semantically contain.
  \item \textit{Retrieval keys}: a set specifying which code or text should be retrieved from the repository when generating the query. Each element contains two fields, \underline{\textit{type}} and \underline{\textit{key}}. The \textit{type} can be either \textit{location} or \textit{keyword}, with the \textit{key} denoting either the path location of the code or text to be retrieved or the query text used for similarity-based retrieval, respectively.
  \item \textit{Query summary}: a concise text summarizing the core content of the query.
\end{itemize}
\end{definition}

\begin{definition}[topic]
A topic is a sequence consisting of multiple query items focusing on the same or closely related themes. Topics are divided into two types: task topics and non-task topics. Query items in the former primarily focus on a sample's information items,
whereas pertain to other plausible repository-level code questions that are largely unrelated to the target function in the latter. 
The primary role of non-task topics is to further introduce conversational noise, making the interaction more reflective of complex real-world scenarios. Each non-task topic 
initially contains a single query item, with all fields left undetermined except for the empty \textit{contained information} set.
\end{definition}

\begin{definition}[query outline]
A query outline is a sequence composed of multiple diverse topics and includes all information items of a sample group, corresponding one-to-one with a conversation.
\end{definition}

First, we partition all filtered samples into multiple sample groups. 
To balance realism with diversity, we limit the size $k$ of sample group to between 1 and 4, and require that all samples within the same group originate from the same repository.
Then, for each sample within a group, we construct several topics based on its information item dependency graph. This process involves three main steps:

\begin{enumerate}[leftmargin=*]
    \item \textbf{Dependency Graph Serialization}: We employ a topological sorting algorithm to serialize the sample's information item dependency graph into an information item unit sequence, denoted as $\{U(I_n)\}_{n=1}^N = (U(I_1), \ldots, U(I_N))$. If multiple source nodes exist during sorting, one is chosen at random. This ensures that the serialized order of the information item units respects causality while maximizing diversity. 
    
    \item \textbf{Information Item Dispersion}: All information item units in the sequence $\{U(I_n)\}_{n=1}^N$ are dispersed into a query item sequence $\{Q_r\}_{r=1}^R$, where the order of the query items reflects the temporal order of queries in conversation. 
    Dispersion follows two primary rules: \ding{182} For information items within the same unit, distracting information items precede the ground-truth one, and each information item must reside in a different query item. 
    For example, if $U(I)$ contains 2 distracting items ($\tilde{I}_1$ and $\tilde{I}_2$) and the query sequence has length $R = 4$,
    one possible distribution is placing $I$, $\tilde{I}_1$, and $\tilde{I}_2$ in the 4th, 2nd, and 1st query items, respectively. This aligns with typical iterative requirement scenarios, where a requirement evolves across multiple turns and its final version appears at the end.
    \ding{183} If an information item unit $U(I_1)$ precedes another unit $U(I_2)$ in the sequence $\{U(I_n)\}_{n=1}^N$, the first information item in the distribution of $U(I_1)$ must appear no later than the first item of $U(I_2)$.
    Information items are then added to the \textit{contained information} set of their assigned query items. If an information item contains code \textit{locations}, it will be also converted 
    added to the \textit{retrieval keys} set.

    \item \textbf{Query Item Partitioning}: Given the predetermined number of topics $N_{\text{topics}}$, we randomly select $N_{\text{topics}}-1$ partition points in the sequence $\{U(I_n)\}_{n=1}^N$ to divide it into $N_{\text{topics}}$ subsequences. Each subsequence forms the ``skeleton'' of a topic, so-called because some fields remain undetermined. 
\end{enumerate}

Finally, we construct the skeleton of a query outline based on all topics of a sample group. After previous steps, the $k$ samples in each group yield $k$ topic sequences. 
We merge these sequences using an algorithm similar to $k$-way merging. 
Instead of selecting the minimum head element, a head element is chosen at random. This preserves the intra-sample topic order while evenly interleaving topics from different samples. Subsequently, we randomly insert several non-task topics into the merged sequence, completing the construction of the query outline skeleton.

\subsubsection{Query Outline Population}
\label{sec:query_outline_population}

Finally, we populate the undetermined fields in the skeleton.

\begin{definition}[recap query item]
A recap query item is a special type of query item whose generated user query is a retrospective question about the agent response in the previous turn, such as 
seeking clarification of the response. It cannot appear in the first position of a topic's query item sequence. Except for the \textit{query type}, all other fields are left undetermined. 
\end{definition}

In the query outline skeleton obtained in the previous step, many fields of query items remain undermined, such as the \textit{query summary} of all query items, and the \textit{query type} and \textit{retrieval keys} of query items in non-task topics. To this end, we assemble the query outline into a prompt, and invoke an LLM to perform population. We define a set of validation rules to ensure the validity of the populated content. Afterward, we randomly insert several recap query items into the outline to further approximate real-world conversation. To ensure that the query outline yields sufficiently long context, while remaining diverse and consistent with typical development practices, we limit the total number of query items $l$ in an outline to $[30, 70)$. Our subsequent experiments show that this length of conversation typically results in conversational contexts reaching up to 64K\textasciitilde256K tokens, which 
satisfies the requirement for evaluating long-horizon context management.
 
\subsection{\ourbench{} Benchmark}

Based on the query outlines generated in \S \ref{sec:automated_construction_of_locoeval}, we construct \ourbench{}, a benchmark for context management in repository-oriented long-horizon conversation. 
It comprises three types of evaluation tasks (\S \ref{sec:task_types_and_evaluation_metrics}) and a dynamic evaluation framework (\S \ref{sec:evaluation_framework}), and is divided into two subsets according to different information distribution patterns (\S \ref{sec:distribution_patterns}), with each subset containing 64 query outlines and 384 evaluation tasks (\S \ref{sec:benchmark_components}).

\subsubsection{Distribution Patterns}
\label{sec:distribution_patterns}

Under different preset values of $N_{\text{topics}}$, we construct two benchmark subsets: a \textbf{\textit{single-hop}} subset ($N_{\text{topics}}$ set to 1) and a \textbf{\textit{multi-hop}} subset ($N_{\text{topics}}$ is randomly chosen between 2 and 3), together forming diverse evaluation scenarios. In the single-hop subset, query items corresponding to the same sample are distributed within a single topic in the query outline, focusing on evaluating an agent's ability to locate and utilize concentrated information. In contrast, in the multi-hop subset, query items are distributed across topics, placing greater emphasis on the agent's capability to identify, track, and integrate dispersed information across the conversation. 

\subsubsection{Tasks and Evaluation Metrics}
\label{sec:task_types_and_evaluation_metrics}

To comprehensively evaluate an agent's long-horizon context management capabilities, we design three types of downstream tasks for each query outline:

\begin{enumerate}[leftmargin=*]
  \item \textbf{Topic Awareness Task}: This is the most basic task, in which the agent is required to output a set of brief descriptions of all topics appearing in the entire conversation. This task assesses the agent's ability to comprehensively understand and summarize the conversation history, which underpins accurate identification of key information over long-horizon interactions. Each query outline is associated with one topic awareness task.
  
  \item \label{item:information_item_extraction_task} \textbf{Information Item Extraction Task}: This task serves as a prerequisite for Task (\ref{item:function_generation_task}). Given the name of a sample's target function, the agent is asked to 
  output all ground-truth information items of this sample according to the conversation.
  This task evaluates the agent's information identification and extraction capability, which is critical for correctly following user requirements and producing accurate responses. Each outline contains $k$ (the size of its corresponding sample group) tasks of this type, with each task corresponding one-to-one to a sample.

  \item \label{item:function_generation_task} \textbf{Function Generation Task}: This is the most important and core task, providing the most comprehensive reflection of context management performance. Given the name of a sample's target function, the agent is asked to generate its complete implementation based on 
  the conversation and the repository. This task examines the agent's ability to integrate conversational context with repository knowledge and supports unbiased assessment through execution-based evaluation. Its IO closely matches real-world repository-level code generation scenarios; therefore, successful implementation best reflects the agent's overall repository-oriented long-horizon conversational context management capability. Within each outline, the number and correspondence pattern of this task are consistent with those of Task (\ref{item:information_item_extraction_task}).
\end{enumerate}

Each task is provided in the form of a predefined natural language query. Correspondingly, for the first two tasks, we use \textbf{F1 score} as the evaluation metric, computed as follows:
\begin{equation}
    \text{Precision} = \frac{|\text{Pred} \cap \text{Gt}|}{|\text{Pred}|}, \quad
    \text{Recall} = \frac{|\text{Gt} \cap \text{Pred}|}{|\text{Gt}|}, \quad
    \text{F1} = \frac{2 \times \text{Precision} \times \text{Recall}}{\text{Precision} + \text{Recall}}
\end{equation}
Here, \text{Pred} and \text{Gt} denote the sets of topics or information items predicted by the agent and the corresponding ground truth, respectively. Their intersection is determined using an LLM-as-a-Judge approach \cite{zheng2023judging}, where matching elements are selected from the left operand of the intersection operation. 
For the function generation task, we use test-case-based \textbf{Pass@k} \cite{chen2021evaluating} as the evaluation metric, reusing test suites provided by DevEval.
Additionally, we also record agents' total LLM \textbf{Token} consumption during evaluation to measure inference cost.

\subsubsection{Evaluation Framework}
\label{sec:evaluation_framework}

\begin{figure}[t]
    \centering
    \setlength{\abovecaptionskip}{0.2cm}
    \includegraphics[width=0.9\linewidth]{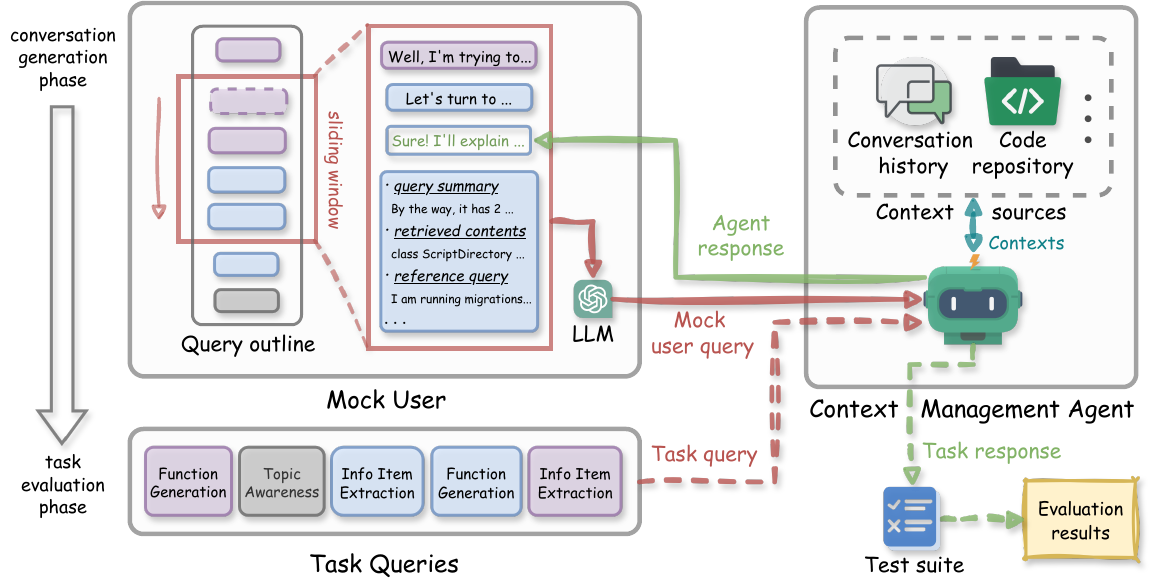}
    \caption{Overview of the evaluation framework of \ourbench{}.}
    \vspace{-0.3cm}
    \label{fig:evaluation_framework}
\end{figure}

To evaluate context management agents on \ourbench{}, we propose a dynamic interactive evaluation framework. An overview of this framework is shown in Figure \ref{fig:evaluation_framework}.
For each sample, the evaluation consists of two phases: \textbf{conversation generation} and \textbf{task evaluation}. In the first phase, to dynamically generate the full conversation and convey information to the agent,
a mock user first generates a mock user query based on the query outline and the immediately preceding response, feeds it to the agent to produce the current response, and then continues to generate the next query in the same way.
This interactive process is repeated $l$ times. 
In the second phase, to evaluate context management based on the generated conversation, we provide the task queries to the agent to obtain task responses, and finally compute the evaluation metrics as described in \S \ref{sec:task_types_and_evaluation_metrics}. 
In particular, since task queries are placed at the end of the conversation and logically independent of one another, they should not change the internal state of agents.
Notably, the framework also monitors all LLM invocations and records token consumption. 

For the mock user queries, after completing $t-1$ interaction turns, the query for the $t$-th turn is generated according to the following steps. First, using the $t$-th query item as the window's end, a window of query items of length 10 is extracted from the query outline,
excluding any recap query items unless it is positioned second to last. 
Then, for all query items within the window, their \textit{query summary} fields are retained. In addition, for the last query item which is the one corresponding to the query to be generated currently, some other fields are also retained or added as follows:

\begin{itemize}[leftmargin=*]
  \item Retain the \textit{query type} and \textit{contained information} fields.
  \item Use the contents in the \textit{retrieval keys} set to retrieve code or text from the repository, providing this as a new \textit{retrieved contents} set. 
  \item Retrieve a real-world question from a Stack Overflow dataset\footnote{It consists of 10,380 high-quality questions curated by us from \url{https://archive.org/details/stackexchange_20250630}.
  } that is semantically similar to the \textit{query summary}. 
  One question is randomly selected from the top 10 most similar
  and is provided as a \textit{reference query} field, aiming to make the generated mock user query better reflect human questioning patterns and preferences, improving simulation authenticity.
\end{itemize}
Finally, the agent's response in the $(t-1)$-th turn is inserted between the second to last and last query items in the window to provide direct context for the current query. This window is incorporated into a carefully designed prompt and fed to a LLM to generate the mock user query.

\subsubsection{Benchmark Composition}
\label{sec:benchmark_components}

The ultimately obtained \ourbench{} consists of two subsets, each containing 64 samples. Each sample includes: \ding{182} a query outline, \ding{183} the corresponding code repository, \ding{184} three types of downstream evaluation tasks totaling $2k + 1$ tasks, and \ding{185} an interactive dynamic evaluation framework. To enhance sample diversity while facilitating systematic analysis, we ensure that all samples within each subset are evenly distributed over the Cartesian product of group size $k$ and query outline length $l$. 
Specifically, among the 64 samples, there are 16 samples for each $k \in \{1,2,3,4\}$.
And within each group of 16 samples, exactly 4 samples fall into each interval of $l$
(\textit{i.e.}, $[30,40)$, $[40,50)$, $[50,60)$, and $[60,70)$).
Therefore, the 64 samples collectively contain $16 \times \sum_{k=1}^42k + 1 = 384$ evaluation tasks.

We further compute the average 
$l$ within each interval, as shown in Table~\ref{tab:avg_l}. The results indicate that, for each subset and interval, the average query outline length 
is close to the midpoint of the corresponding interval, demonstrating the effectiveness of the interval partitioning. In addition, we randomly sampled some query outlines for manual quality inspection and found no obvious errors.

\section{Evaluation}

\subsection{Research Questions}

Based on our \ourbench{}, we aim to answer the following research questions:

\begin{itemize}[leftmargin=*]
    \item \textbf{RQ1: Performance of Standalone LLMs}. How capable are mainstream LLMs at native long-horizon conversational context processing on \ourbench{}?
    \item \textbf{RQ2: Performance of Context Management Methods}. How do existing general-purpose context management methods perform on \ourbench{}?
    \item \textbf{RQ3: Adapting to Repository-Oriented Scenarios}. How can general-purpose context management methods be improved to better suit repository-oriented scenarios?
    \item \textbf{RQ4: Impact of Conversation Hyperparameters}. How do various conversation hyperparameters affect the performance of context management agents on \ourbench{}?
\end{itemize}

\begin{table}[t]
\centering
\small
\setlength{\abovecaptionskip}{0.1cm}
\caption{Average Query Outline Length $l$ in Each Interval.}
\label{tab:avg_l}
\begin{tabular}{lccccc}
\toprule
\textbf{Subset} & \textbf{[30,40)} & \textbf{[40,50)} & \textbf{[50,60)} & \textbf{[60,70)} & \textbf{All} \\
\midrule
\textit{Single-hop} & 33.19 & 44.88 & 54.13 & 64.63 & 49.20 \\
\textit{Multi-hop} & 37.44 & 45.75 & 55.00 & 64.38 & 50.64 \\
\bottomrule
\end{tabular}
\vspace{-0.3cm}
\end{table}

\subsection{Baselines and Evaluation Details}

We select a standalone LLM method without context management and four representative context management methods as baselines, which are wrapped as context management agents:

\begin{itemize}[leftmargin=*]
    \item \textbf{\textit{Full}} denotes a standalone LLM that truncate conversation history at the turn level only when the context exceeds the window limit, without any additional processing.
    \item \textbf{\textit{Vanilla RAG}} retrieves the top-$k$ most similar queries to the current user query from the current conversation history, 
    and places the corresponding interaction turns before the current query as context. Balancing performance and token overhead, we set $k = 5$.
    \item \textbf{MemGPT} \cite{packer2023memgpt} is an OS-inspired context management system that uses hierarchical memory to extend an LLM’s context window, enabling long-term, multi-session conversations.
    \item \textbf{LD-Agent} \cite{li2025hello} is a long-horizon conversational agent framework that combines event-based memory with persona modeling, enabling the agent to maintain consistency during interactions.
    \item \textbf{Mem0} \cite{chhikara2025mem0} is a production-oriented long-horizon memory architecture that dynamically extracts, integrates, and retrieves key information from conversations. 
\end{itemize}

The hyperparameters used in the last three memory systems are kept identical to those reported in their original papers.
Given that the baselines are designed for general-purpose conversations, we introduce a unified RAG adaptation strategy to make them preliminarily applicable to the repository-oriented scenarios emphasized by \ourbench{}.
Specifically, after receiving each user query in a turn, each agent retrieves the top 10 most similar functions to the query from the repository
and supplement them to the context as background knowledge. 

In addition to the above baselines, to understand the approximate upper and lower bounds of context management agents and compare it with their actual performance, we introduce the \textbf{\textit{Oracle}} and \textbf{\textit{Empty}} baselines. 
\textit{Empty} refers to a baseline that provides no conversational context other than the current user query and the top 10 similar functions. Therefore, it can be regarded as the \textbf{lower bound}. 
As for \textit{Oracle}, instead of relying on dynamically generated user queries, this baseline directly uses the ground-truth information items obtained as described in \S \ref{sec:information_items_extraction_and_mutation}, along with the contents of code elements associated with these information items, as context, thus avoiding information bias or loss during context management. Benefiting from this omniscient perspective, \textit{Oracle} essentially represents the \textbf{upper bound};
in particular, for the topic awareness and information item extraction tasks that do not require additional reasoning, evidently, it can directly achieve the maximum F1 score of 1. However, due to the potential incomplete coverage of ground-truth information items and possible reasoning errors during generation, it cannot achieve perfect correctness in the function generation task. Consequently, if more relevant information could be retrieved from the repository or more advanced generation workflows were employed, the agent performance could potentially surpass that of the \textit{Oracle} on this task.

We further define two extended metrics to more reasonably characterize the context management capability.
The first is the \textbf{Compression Ratio}, which reflects to what extent an agent can reduce context consumption compared to a setting with no context management. A higher value indicates relatively lower context overhead. For an agent $A$, it is computed as:
\begin{equation}
\label{form:cr}
    \text{Compression Ratio} = \frac{\text{Tokens of } A}{\text{Tokens of \textit{Full}}}
\end{equation}
The second is the \textbf{Normalized F1/Pass@k Score}. For an agent $A$, it is computed as:
\begin{equation}
\label{form:ns}
    \text{Normalized Score} = \frac{\text{Score of } A - \text{Score of \textit{Empty}}}{\text{Score of \textit{Oracle}} - \text{Score of \textit{Empty}}} \times 100\%
\end{equation}
This reflects the extent to which an agent can close the gap between the theoretical upper bound (\textit{Oracle}) and lower bound (\textit{Empty}) of repository-oriented context management performance. A higher value indicates better effectiveness. Compared with the original score, the normalized score can largely factor out the influence of confounding factors such as task difficulty and potential data contamination \cite{sainz2023nlp, balloccu2024leak}, 
thereby more accurately reflecting the actual performance.
In Formula (\ref{form:cr}) and (\ref{form:ns}), \textit{Full}, \textit{Empty}, and \textit{Oracle} are all evaluated under the same experimental setup as agent $A$.

For each baseline, we conduct experiments using 3 advanced backbone LLMs,
including \textbf{GPT-5 mini}
\cite{openai2025gpt5mini}, \textbf{DeepSeek-V3.2} \cite{liu2025deepseek32}, and \textbf{Qwen3-235B-A22B}\footnote{We use its \texttt{qwen3-235b-a22b-instruct-2507} snapshot.} \cite{yang2025qwen3}, whose maximum context window token lengths are 400K, 128K, and 256K, respectively. 
Throughout the evaluation, all embedding models utilize \textit{all-MiniLM-L6-v2} \cite{wang2020minilm}. The LLM used for generating mock user queries and 
LLM-as-a-Judge evaluating 
is \textit{Gemini 2.5 Flash}
\cite{comanici2025gemini}, with the former running without CoT enabled. To reduce variability caused by probabilistic sampling, we set the temperature of all LLMs to 0.

\section{Results and Analysis}

\subsection{RQ1: Performance of Standalone LLMs}

To evaluate the native long-horizon conversational context processing capability of standalone LLMs and to empirically highlight the necessity of context management, we first apply \ourbench{} to the \textit{Full} baseline. 
Table \ref{tab:rq1_result} reports the performance of different standalone LLMs across all three tasks, together with their token consumption. The results show that, within our evaluation scope, existing LLMs generally suffer from clear performance bottlenecks when confronted with ultra-long contexts. Among them, DeepSeek-V3.2 performs the worst, followed by Qwen3-235B-A22B, while GPT-5 mini performs the best with only one exception.
This ranking is consistent with their maximum context window lengths, indicating that larger context windows indeed often correspond to stronger long-context capabilities. Nevertheless, even for the best-performing GPT-5 mini, the normalized scores on both the information item extraction and function generation tasks do not exceed 50\%, implying a performance loss of more than half.
Moreover, the average token consumption of all LLMs exceeds 3M tokens; even for the lowest-priced DeepSeek-V3.2 API rate of \$0.28/1M input tokens, each conversation would still cost at least around \$1. Taken together, these results clearly demonstrate the multiple challenges faced by standalone LLMs when handling repository-oriented long-horizon conversations. 

A horizontal comparison across tasks further reveals that scores on the topic awareness task are consistently and substantially higher than those on the other two tasks. This is likely because this task only requires the LLM to grasp high-level themes rather than fine-grained details, suggesting that the loss of fine-grained information is a more prominent issue than topic drift in long-horizon conversations. 
Interestingly, although there is a general performance downward trend 
from the information item extraction task to the function generation task, GPT-5 mini exhibits an anomalous and consistent improvement.
This may be attributed to its strong coding robustness, which enables it to more accurately identify relevant information from noisy contexts to support code generation. Moreover, although the single-hop and multi-hop subsets have essentially identical average conversation turns (see Table \ref{tab:avg_l}), 
LLMs exhibit a somewhat counterintuitive trend of consistently higher normalized scores
on the seemingly more complex multi-hop subset than on the single-hop subset. We conjecture that, under the dispersed information distribution pattern of the multi-hop subset, information is distributed more evenly across the whole interaction, thereby improving the LLMs' fault tolerance in accurately identifying and leveraging relevant information.

\begin{table}[t]
\centering
\setlength{\abovecaptionskip}{0.1cm}
\caption{Performance of different standalone LLMs on \ourbench{}. The metrics for the first two tasks are F1 scores, while the metric for the third task is Pass@1. The percentages in parentheses indicate the normalized scores. \textit{Avg. \#Token} 
is computed as the total prompt tokens of \textit{Full} divided by the number of samples.}
\label{tab:rq1_result}
\resizebox{\linewidth}{!}{
\begin{tabular}{lcccccccccc}
\toprule
\multirow{2}{*}{\textbf{Model}} & \multicolumn{3}{c}{\textbf{Topic Awareness}} & \multicolumn{3}{c}{\textbf{Info Item Extraction}} &\multicolumn{3}{c}{\textbf{Function Generation}} & \textbf{Avg.} \\
\cmidrule(lr){2-4} \cmidrule(lr){5-7} \cmidrule(lr){8-10}
& \textit{Oracle} & \textit{Empty} & \cellcolor{LightGray} \textit{Full} & \textit{Oracle} & \textit{Empty} & \cellcolor{LightGray} \textit{Full} & \textit{Oracle} & \textit{Empty} & \cellcolor{LightGray} \textit{Full} & \textbf{\# Token} \\
\midrule
\multicolumn{11}{c}{\textbf{\textit{Single-hop}}} \\
\cmidrule(lr{0.6em}){1-11}
GPT-5 mini & 1.000 & 0.187 & \textbf{\cellcolor{LightGray} 0.871~{\color{DeepGray} (84.18\%)}} & 1.000 & 0.029 & \cellcolor{LightGray} \textbf{0.435~{\color{DeepGray} (41.83\%)}} & 63.75 & 6.88 & \cellcolor{LightGray} \textbf{35.00~{\color{DeepGray} (49.45\%)}} & 3.27M \\
Deepseek-V3.2 & 1.000 & 0.266 & \cellcolor{LightGray} 0.650~{\color{DeepGray} (52.27\%)} & 1.000 & 0.062 & \cellcolor{LightGray} 0.356~{\color{DeepGray} (31.40\%)} & 55.00 & 6.25 & \cellcolor{LightGray} 8.75~{\color{DeepGray} (5.13\%)} & 4.44M \\
Qwen3-235B-A22B & 1.000 & 0.025 & \cellcolor{LightGray} 0.630~{\color{DeepGray} (62.00\%)} & 1.000 & 0.009 & \cellcolor{LightGray} 0.365~{\color{DeepGray} (35.93\%)} & 51.88 & 3.12 & \cellcolor{LightGray} 13.75~{\color{DeepGray} (21.79\%)} & 4.07M \\
\cmidrule(lr{0.6em}){1-11}
\multicolumn{11}{c}{\textbf{\textit{Multi-hop}}} \\
\cmidrule(lr{0.6em}){1-11}
GPT-5 mini & 1.000 & 0.178 & \cellcolor{LightGray} \textbf{0.887~{\color{DeepGray} (86.22\%)}} & 1.000 & 0.032 & \cellcolor{LightGray} 0.385~{\color{DeepGray} (36.46\%)} & 63.75 & 7.50 & \cellcolor{LightGray} \textbf{35.62~{\color{DeepGray} (50.00\%)}} & 3.17M \\
Deepseek-V3.2 & 1.000 & 0.294 & \cellcolor{LightGray} 0.738~{\color{DeepGray} (62.91\%)} & 1.000 & 0.073 & \cellcolor{LightGray} 0.416~{\color{DeepGray} (37.07\%)} & 58.12 & 5.00 & \cellcolor{LightGray} 11.88~{\color{DeepGray} (12.94\%)} & 4.17M \\
Qwen3-235B-A22B & 1.000 & 0.018 & \cellcolor{LightGray} 0.766~{\color{DeepGray} (76.22\%)} & 1.000 & 0.012 & \cellcolor{LightGray} \textbf{0.421~{\color{DeepGray} (41.43\%)}} & 52.50 & 3.12 & \cellcolor{LightGray} 18.75~{\color{DeepGray} (31.65\%)} & 3.81M \\
\bottomrule
\end{tabular}
}
\vspace{-0.3cm}
\end{table}

\vspace{3mm}
\begin{custommdframed}
\textbf{\textit{Answer to RQ1:}} Standalone LLMs exhibit substantial performance degradation when handling repository-oriented long-horizon conversations, losing over half of their effective performance on fine-grained tasks while incurring prohibitive token costs. Although larger context windows could alleviate this issue to some extent and dispersed information distributions are more robust, these results clearly demonstrate that native long-horizon conversational context capacity alone is insufficient, underscoring the necessity of explicit context management.
\end{custommdframed}
\vspace{0mm}

\subsection{RQ2: Performance of Context Management Agents}

We further evaluate all context management agents on \ourbench{}, with the results summarized in Table \ref{tab:rq2_result}. 
Given that \ding{172} the agents' performance on the function generation task could largely reflect that on the information item extraction task, \ding{173} the results in RQ1 show that performance on the topic awareness task is significantly higher than on the other tasks, indicating that it is not a bottleneck, and \ding{174} both topic awareness and information item extraction are non-core and intermediate-stage tasks rather than user-facing outcomes,
we therefore report results only for the core function generation task in the subsequent RQs.

The results in Table \ref{tab:rq2_result} show that, compared to \textit{Full}, the best-performing one of the other four agents consistently achieves performance improvements of varying magnitudes. The largest improvement is observed when DeepSeek-V3.2 is used as the backbone LLM, with an absolute gain of approximately 34 percentage points in the normalized score. At the same time, context consumption is significantly reduced, achieving at least a $5\times$ compression ratio except MemGPT.
These demonstrate the effectiveness of existing context management methods on \ourbench{} to a certain extent.
However, a prominent observation is that the simple \textit{Vanilla RAG} achieves the best performance in most cases, whereas the carefully designed memory systems generally underperform.
Moreover, the absolute performance of \textit{Vanilla RAG} itself is not particularly strong, especially when paired with the weaker backbone LLMs DeepSeek-V3.2 and Qwen3-235B-A22B. Together, these indicate that existing general-purpose context management methods have substantial room for improvement in repository-oriented scenarios. How to effectively integrate and leverage both conversational and repository information within memory systems thus emerges as a key research challenge.

In addition, Some memory systems exhibit large performance variations across different backbone LLMs, and even yield negative normalized scores in certain cases, which may be attributed to suboptimal designs of their internal workflows, making them fail to adapt well to specific external configurations when the task scenario changes.
All three memory systems consume significantly more context on GPT-5 mini than on other backbone LLMs. This is likely because GPT-5 mini tends to produce longer outputs when performing tool calls, an operation usually required by memory systems, thereby increasing the context length in subsequent steps.

\begin{table}[t]
\centering
\setlength{\abovecaptionskip}{0.1cm}
\caption{Performance across different context management agents and backbone LLMs on \ourbench{}.}
\label{tab:rq2_result}
\resizebox{\linewidth}{!}{
\begin{tabular}{lcccccc}
\toprule
\multirow{2}{*}{\textbf{Method}} & \multicolumn{2}{c}{\textbf{\textit{GPT-5 mini}}} & \multicolumn{2}{c}{\textbf{\textit{Deepseek-V3.2}}} & \multicolumn{2}{c}{\textbf{\textit{Qwen3-235B-A22B}}} \\
& \textit{pass@1} & \textit{compression ratio} & \textit{pass@1} & \textit{compression ratio} & \textit{pass@1} & \textit{compression ratio}  \\
\midrule
\multicolumn{7}{c}{\textbf{\textit{Single-hop}}} \\
\cmidrule(lr{0.6em}){1-7}
\textbf{\textit{Oracle}} & 63.75~{\color{DeepGray} (100.0\%)} & - & 55.00~{\color{DeepGray} (100.0\%)} & - & 51.88~{\color{DeepGray} (100.0\%)} & - \\
\textbf{\textit{Empty}} & 6.88~{\color{DeepGray} (0.0\%)} & - & 6.25~{\color{DeepGray} (0.0\%)} & - & 3.12~{\color{DeepGray} (0.0\%)} & - \\
\arrayrulecolor{DeepGray}
\cmidrule(lr{0.6em}){1-7}
\arrayrulecolor{black}
\rowcolor{LightGray} \textit{Full} & 35.00~{\color{DeepGray} (49.45\%)} & $1.0 \times$ & 8.75~{\color{DeepGray} (5.13\%)} & $1.0 \times$ & 13.75~{\color{DeepGray} (21.79\%)} & $1.0 \times$ \\
\textit{Vanilla RAG} & \textbf{40.62~{\color{DeepGray} (59.34\%)}} & $5.7 \times$ & \textbf{25.62~{\color{DeepGray} (39.74\%)}} & $5.1 \times$ & 16.88~{\color{DeepGray} (28.21\%)} & $5.6 \times$ \\
MemGPT & 6.25~{\color{DeepGray} (-1.10\%)} & $1.0 \times$ & 8.75~{\color{DeepGray} (5.13\%)} & $1.2 \times$ & \textbf{18.12~{\color{DeepGray} (30.77\%)}} & $1.3 \times$ \\
LD-Agent & 8.75~{\color{DeepGray} (3.30\%)} & $9.8 \times$ & 6.88~{\color{DeepGray} (1.28\%)} & $15.0 \times$ & 4.38~{\color{DeepGray} (2.56\%)} & $13.1 \times$ \\
Mem0 & 30.62~{\color{DeepGray} (41.76\%)} & $5.3 \times$ & 5.62~{\color{DeepGray} (-1.28\%)} & $15.6 \times$ & 5.00~{\color{DeepGray} (3.85\%)} & $15.4 \times$ \\
\cmidrule(lr{0.6em}){1-7}
\multicolumn{7}{c}{\textbf{\textit{Multi-hop}}} \\
\cmidrule(lr{0.6em}){1-7}
\textbf{\textit{Oracle}} & 63.75~{\color{DeepGray} (100.0\%)} & - & 58.12~{\color{DeepGray} (100.0\%)} & - & 52.5~{\color{DeepGray} (100.0\%)} & - \\
\textbf{\textit{Empty}} & 7.50~{\color{DeepGray} (0.0\%)} & - & 5.00~{\color{DeepGray} (0.0\%)} & - & 3.12~{\color{DeepGray} (0.0\%)} & - \\
\arrayrulecolor{DeepGray}
\cmidrule(lr{0.6em}){1-7}
\arrayrulecolor{black}
\rowcolor{LightGray} \textit{Full} & 35.62~{\color{DeepGray} (50.00\%)} & $1.0 \times$ & 11.88~{\color{DeepGray} (12.94\%)} & $1.0 \times$ & 18.75~{\color{DeepGray} (31.65\%)} & $1.0 \times$ \\
\textit{Vanilla RAG} & \textbf{36.88~{\color{DeepGray} (52.22\%)}} & $6.4 \times$ & \textbf{26.88~{\color{DeepGray} (41.18\%)}} & $6.1 \times$ & 18.12~{\color{DeepGray} (30.38\%)} & $5.8 \times$ \\
MemGPT & 12.50~{\color{DeepGray} (8.89\%)} & $1.0 \times$ & 14.38~{\color{DeepGray} (17.65\%)} & $1.1 \times$ & \textbf{21.25~{\color{DeepGray} (36.71\%)}} & $1.1 \times$ \\
LD-Agent & 8.12~{\color{DeepGray} (1.11\%)} & $10.4 \times$ & 7.50~{\color{DeepGray} (4.71\%)} & $15.5 \times$ & 6.25~{\color{DeepGray} (6.33\%)} & $12.9 \times$ \\
Mem0 & 30.62~{\color{DeepGray} (41.11\%)} & $5.6 \times$ & 7.50~{\color{DeepGray} (4.71\%)} & $15.8 \times$ & 9.38~{\color{DeepGray} (12.66\%)} & $15.8 \times$ \\
\bottomrule
\end{tabular}
}
\vspace{-0.4cm}
\end{table}

\vspace{3mm}
\begin{custommdframed}
\textbf{\textit{Answer to RQ2:}} Overall, the best context management agent consistently improve performance over \textit{Full} on function generation task while substantially reducing context consumption, demonstrating the effectiveness of context management on \ourbench{}. However, the relatively strong performance of the simple \textit{Vanilla RAG} and the underperformance of more complex memory systems reveal that existing general-purpose context management methods remain poorly adapted to repository-oriented long-horizon conversations, leaving significant room for improvement.
\end{custommdframed}
\vspace{0mm}

\subsection{RQ3: Adapting to Repository-Oriented Scenarios}

\begin{figure}[t]
    \centering
    \setlength{\abovecaptionskip}{0.1cm}
    \includegraphics[width=0.85\linewidth]{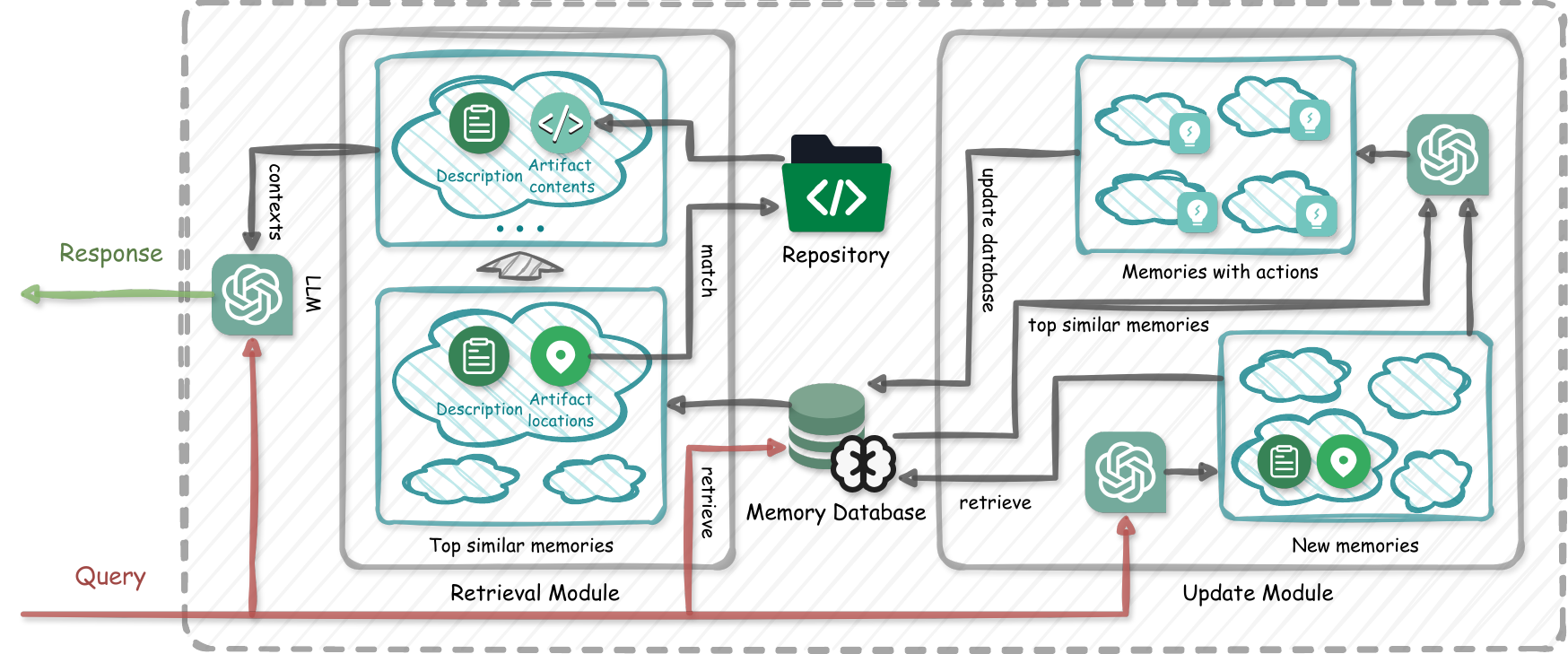}
    \caption{Agent workflow of our improved Mem0$^\mathcal{R}$.}
    \vspace{-0.2cm}
    \label{fig:mem0r}
\end{figure}

The results in RQ2 indicate that existing methods especially memory systems designed for general-purpose conversations struggle to adapt to the repository-oriented scenarios targeted by \ourbench{}, leading to generally poor performance. To explore preliminary directions for addressing this issue, we propose Mem0$^\mathcal{R}$, an improved variant of Mem0 \cite{chhikara2025mem0} specifically tailored for repository-oriented context management, 
as illustrated in Figure \ref{fig:mem0r}. 
Mem0 maintains a memory database to store conversational memories, and it consists of three main phases: \textit{memory extraction}, \textit{memory integration}, and \textit{memory retrieval}. After each conversation turn, Mem0 extracts a set of textual memories from the current interaction, compares them with similar memories in the database, and decides whether to add, update, or delete, thereby integrating new memories into the database. When a new user query arrives, Mem0 retrieves the most similar memories from the database using the query and incorporates them into the LLM context as background knowledge.

\begin{table}[t]
\centering
\setlength{\abovecaptionskip}{0.1cm}
\caption{Comparison of the performance between our improved Mem0$^\mathcal{R}$ and the best baseline, \textit{Vanilla RAG}.
}
\label{tab:rq3_result}
\resizebox{\linewidth}{!}{
\begin{tabular}{lcccccc}
\toprule
\multirow{2}{*}{\textbf{Method}} & \multicolumn{2}{c}{\textbf{\textit{GPT-5 mini}}} & \multicolumn{2}{c}{\textbf{\textit{Deepseek-V3.2}}} & \multicolumn{2}{c}{\textbf{\textit{Qwen3-235B-A22B}}} \\
& \textit{pass@1} & \textit{compression ratio} & \textit{pass@1} & \textit{compression ratio} & \textit{pass@1} & \textit{compression ratio}  \\
\midrule
\multicolumn{7}{c}{\textbf{\textit{Single-hop}}} \\
\cmidrule(lr{0.6em}){1-7}
\textbf{\textit{Oracle}} & 63.75~{\color{DeepGray} (100.0\%)} & - & 55.00~{\color{DeepGray} (100.0\%)} & - & 51.88~{\color{DeepGray} (100.0\%)} & - \\
\textbf{\textit{Empty}} & 6.88~{\color{DeepGray} (0.0\%)} & - & 6.25~{\color{DeepGray} (0.0\%)} & - & 3.12~{\color{DeepGray} (0.0\%)} & - \\
\arrayrulecolor{DeepGray}
\cmidrule(lr{0.6em}){1-7}
\arrayrulecolor{black}
\textit{Vanilla RAG} & \textbf{40.62~{\color{DeepGray} (59.34\%)}} & $5.7 \times$ & 25.62~{\color{DeepGray} (39.74\%)} & $5.1 \times$ & 16.88~{\color{DeepGray} (28.21\%)} & $5.6 \times$ \\
\rowcolor{LightGray} \textbf{Mem0$^\mathcal{R}$} & 37.50~{\color{DeepGray} (53.85\%)} & $2.8 \times$ & \textbf{31.88~{\color{DeepGray} (52.56\%)}} & $5.1 \times$ & \textbf{24.38~{\color{DeepGray} (43.59\%)}} & $6.1 \times$ \\
\cmidrule(lr{0.6em}){1-7}
\multicolumn{7}{c}{\textbf{\textit{Multi-hop}}} \\
\cmidrule(lr{0.6em}){1-7}
\textbf{\textit{Oracle}} & 63.75~{\color{DeepGray} (100.0\%)} & - & 58.12~{\color{DeepGray} (100\%)} & - & 52.5~{\color{DeepGray} (100.0\%)} & - \\
\textbf{\textit{Empty}} & 7.50~{\color{DeepGray} (0.0\%)} & - & 5.00~{\color{DeepGray} (0\%)} & - & 3.12~{\color{DeepGray} (0.0\%)} & - \\
\arrayrulecolor{DeepGray}
\cmidrule(lr{0.6em}){1-7}
\arrayrulecolor{black}
\textit{Vanilla RAG} & 36.88~{\color{DeepGray} (52.22\%)} & $6.4 \times$ & 26.88~{\color{DeepGray} (41.18\%)} & $6.1 \times$ & 18.12~{\color{DeepGray} (30.38\%)} & $5.8 \times$ \\
\rowcolor{LightGray} \textbf{Mem0$^\mathcal{R}$} & \textbf{42.50~{\color{DeepGray} (62.22\%)}} & $3.2 \times$ & \textbf{37.50~{\color{DeepGray} (61.18\%)}} & $5.4 \times$ & \textbf{25.62~{\color{DeepGray} (45.57\%)}} & $6.4 \times$ \\
\bottomrule
\end{tabular}
}
\vspace{-0.3cm}
\end{table}

In order to integrate both conversational and repository information, the key modification introduced by Mem0$^\mathcal{R}$ is that memories are no longer purely textual. Instead, each memory is a composite structure consisting of a textual description and the path locations of associated repository artifacts (\textit{e.g.}, documents or functions). During retrieval, these locations are used to fetch the corresponding content from the repository. This design enables memories to capture not only salient conversational information but also explicit links to concrete repository artifacts, thereby supporting more precise and context-aware repository retrieval. Notably, since the locations extracted from queries may be incomplete, when an exact match cannot be found we fall back to fuzzy matching: we first filter candidate locations based on the final component of the location, and then select the one most similar to the full target location.
We evaluated Mem0$^\mathcal{R}$ on \ourbench{} and compared it with the best baseline \textit{Vanilla RAG}. The results in Table \ref{tab:rq3_result} show that Mem0$^\mathcal{R}$ outperforms \textit{Vanilla RAG} under almost all settings, while maintaining comparable compression rates, except for the special case of GPT-5 mini, which we have discussed in RQ2. This not only demonstrates the effectiveness of Mem0$^\mathcal{R}$ but also corroborates the conclusion in RQ2 that memory systems hold considerable potential for improvement in repository-oriented long-horizon conversations.

\vspace{3mm}
\begin{custommdframed}
\textbf{\textit{Answer to RQ3:}} To better adapt general-purpose context management methods for repository-oriented scenarios, we made a preliminary improvement to integrate conversational memories with explicit links to relevant repository artifacts, enabling more accurate retrieval of repository. Our evaluation demonstrates that this approach, named Mem0$^\mathcal{R}$, enhances performance while maintaining efficiency, highlighting the promising potential of memory systems in this scenario.
\end{custommdframed}
\vspace{0mm}

\subsection{RQ4: Impact of Conversation Hyperparameters}

\begin{figure}[t]
    \centering
    \setlength{\abovecaptionskip}{0.1cm}
    \includegraphics[width=\linewidth]{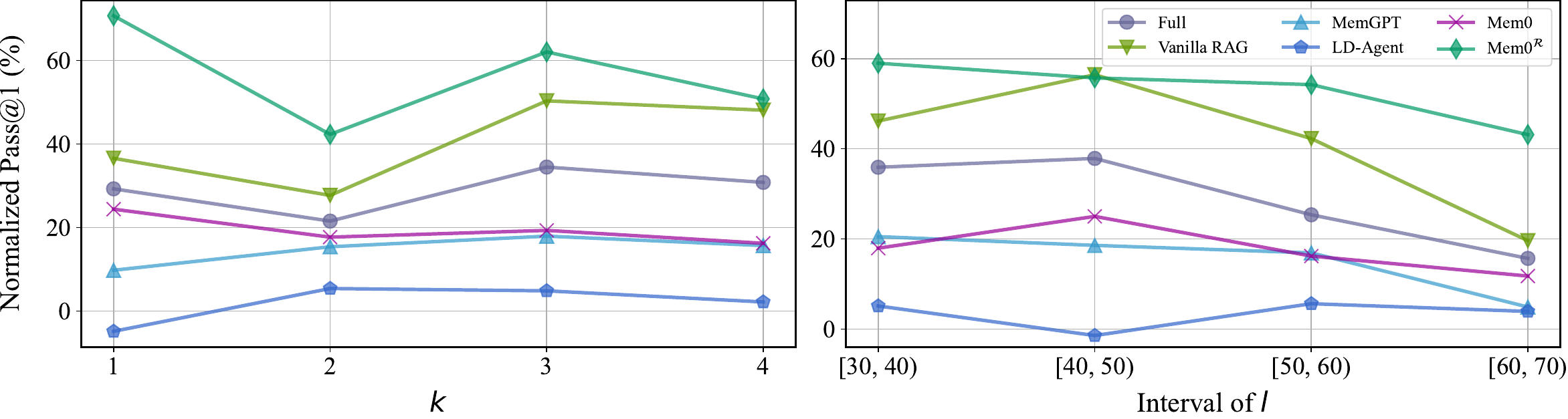}
    \caption{Trends of the normalized pass@1 on the function generation task for different agents, with respect to the number of this task $k$ per sample (left) and the interval of conversation length $l$ per sample (right).}
    \vspace{-0.3cm}
    \label{fig:k_and_l_analysis}
\end{figure}

The performance of context management agents can be influenced by various external factors, in particular by the hyperparameters of the conversation itself. For instance, longer conversations intuitively tend to 
posing greater challenges for agents in accurately managing and retrieving relevant information. Understanding the actual impact of these factors is therefore highly valuable for both researchers and developers. After careful consideration, we focus our analysis on two key hyperparameters: the number of function generation tasks $k$ and the conversation length $l$ (measured in turns) of each sample. These two factors respectively reflect the diversity and the volume of the context. Analyzing them allows us to directly examine the generality and robustness of agents under different conditions. 

As shown in Figure \ref{fig:k_and_l_analysis}, variations in $k$ have no clear overall impact on performance\footnote{The minimum chi-square test $p$-value across all agents is $0.1096 > 0.05$, indicating no statistically significant differences.}. Although some fluctuations exist, there is no evident consistent 
trend, indicating that agents are generally able to accurately acquire relevant information even under diverse and complex contexts, and can effectively handle multi-task development scenarios. In contrast, as $l$ increases, most agents exhibit a clear downward trend in performance\footnote{Except for LD-Agent, the maximum $p$-value across agents is $0.0351 < 0.05$, indicating statistically significant differences.} (at most after a slight initial increase), which aligns with our intuition. 
A possible underlying mechanism for this degradation is that, for \textit{Vanilla RAG}, it stems from the increase in noisy turns, whereas for memory systems it arises from excessive iterations that impair the overall quality of the memory.
This degradation may limit the broad applicability of context management methods and future research therefore should place greater emphasis on enabling methods to maintain performance as conversation grows continuously. Notably, when the interval of $l$ increases from $[30,40)$ to $[60,70)$, the performance of our proposed Mem0$^{\mathcal{R}}$ drops by only 26.85\%. By comparison, 
the two suboptimal agents,
\textit{Vanilla RAG} and \textit{Full}, suffer much larger declines of 57.52\% and 56.30\%, respectively, further demonstrating the robustness of Mem0$^{\mathcal{R}}$. 

\vspace{3mm}
\begin{custommdframed}
\textbf{\textit{Answer to RQ4:}} The number of function generation tasks $k$ has no statistically significant impact on performance, indicating that agents are generally robust to variations in contextual diversity. In contrast, increasing conversation length $l$ significantly degrades performance for most agents, stressing the necessity of studying how to maintain performance under growing interaction turns.
\end{custommdframed}
\vspace{0mm}

\section{Discussion}

\subsection{Adherence to Key Principles}
\label{sec:adherence_to_key_principles}

As the sole medium of interaction between the mock user and agent,
whether the mock user queries adhere to the key principles of correctness, realism, and diversity essentially reflects whether \ourbench{} satisfies these key principles.
Since the queries are generated dynamically during evaluation, perfect adherence cannot be guaranteed in advance. Nevertheless, we adopt a series of strategies to maximize their adherence to these principles.

\subsubsection{Correctness} 

To improve correctness, we employ carefully designed structured prompts for query generation, which appropriately incorporate the query outline and the conversation history. We use \textit{Gemini 2.5 Flash}, one of the most advanced lightweight LLMs currently available, to perform the generation. In addition, we conduct post hoc verification to further ensure correctness. Specifically, We randomly sampled several queries during evaluation for manual inspection and observe no obvious errors, and the fact that Mem0$^\mathcal{R}$ achieves a normalized pass@1 exceeding 60\% further provides indirect evidence of the high correctness of the generated queries.

\subsubsection{Realism and Diversity} 

To improve realism and diversity, our primary strategy is to retrieve similar questions from a real-world question dataset and use them as references during query generation, which clearly improves the realism of the queries, while the inherent diversity of the large-scale dataset naturally enhances query diversity as well. Furthermore, we introduce randomness at multiple phases of the query outline construction pipeline, including probabilistic sampling in LLM invocations. As a result, the query outlines exhibit substantial diversity, which is inherited by the queries. Additionally, employing the query taxonomy derived from an empirical study \cite{peng2025swe} further strengthens the realism of the generated queries.

\subsection{Limitations}

\noindent \textbf{Single programming language.} 
We follow the common practice of existing repository-oriented benchmarks \cite{zhang2023repocoder, li2024deveval, yu2024codereval, ding2023crosscodeeval, li2024evocodebench, jimenez2023swe, li2025longcodeu, zhang2025swe, sunrepofixeval, zhao2025towards, wang2025codeflowbench}, which are typically constructed solely on the most widely used programming language, Python. However, this choice overlooks other languages that are also widely used in repository development (\textit{e.g.}, Java and C++), many of whose features differ substantially from those of Python. We thus plan to incorporate these languages in future work.

\vspace{0.1cm}
\noindent \textbf{Limited evaluation tasks.} 
Considering both evaluation feasibility and difficulty, we use repository-level code generation as the data source and the core evaluation task. Nevertheless, this task cannot fully represent the complexity and diversity encountered in real-world repository development. In future work, we will further explore how to construct context management benchmarks based on tasks such as issue resolution and vulnerability repair.

\vspace{0.1cm}
\noindent \textbf{Limited baselines.} 
Due to the relatively high evaluation cost of long-horizon multi-turn conversations, together with our limited research budget, we are constrained to focus on a small number of representative context management methods and affordable advanced backbone LLMs. However, since both our data and evaluation framework are open-sourced, any researcher can apply their own methods or LLMs to \ourbench{} for evaluation.

\section{Conclusion}

In this work, we introduced \ourbench{}, the first benchmark dedicated to evaluating long-horizon conversational context management in repository-oriented development scenarios. Through extensive experiments, we demonstrated that both standalone LLMs and existing general-purpose context management methods struggle to effectively handle repository-oriented long-horizon conversations.
To address this gap, we further proposed Mem0$^{\mathcal{R}}$, a simple yet effective extension that integrates conversation history with code repository, achieving superior performance and robustness across varying conditions. We hope \ourbench{} and our findings will facilitate future research on more effective and scalable context management for real-world code assistant.

\section*{Data Availability}

Our \ourbench{} and Mem0$^{\mathcal{R}}$ are available at \url{https://anonymous.4open.science/r/LoCoEval}.

\bibliographystyle{ACM-Reference-Format}
\bibliography{refs}

\end{document}